\documentstyle[10pt,epsf,epsfig,hangcaption,xspace,amssymb,amsfonts,amsmath,amsthm,cite,
dp_delphititle,lineno]{dp_delphi}
\makeindex
\def\DpPaperGroup{EP}
\def\DpPaperRef{2002-077}
\def\DpDate{16 October 2002}
\def\DpAuthors{DELPHI Collaboration}
\def\DpSubmit{(Phys. Lett. B552 (2003) 127)}
\def\DpTitle{Search for Doubly Charged Higgs Bosons at LEP2}
\def\DpComment{ }
\def\DpEMail{ }

\newcommand{\gev}{{\ifmmode \mbox{Ge\kern-0.2exV}
\else Ge\kern-0.2exV\nolinebreak\fi}}
\newcommand{\mev}{{\ifmmode \mbox{Me\kern-0.2exV}
\else Me\kern-0.2exV\nolinebreak\fi}}

\begin{document}
\makeatletter
\newcount\@tempcntc
\def\@citex[#1]#2{\if@filesw\immediate\write\@auxout{\string\citation{#2}}\fi
  \@tempcnta\z@\@tempcntb\m@ne\def\@citea{}\@cite{\@for\@citeb:=#2\do
    {\@ifundefined
       {b@\@citeb}{\@citeo\@tempcntb\m@ne\@citea\def\@citea{,}{\bf ?}\@warning
       {Citation `\@citeb' on page \thepage \space undefined}}%
    {\setbox\z@\hbox{\global\@tempcntc0\csname b@\@citeb\endcsname\relax}%
     \ifnum\@tempcntc=\z@ \@citeo\@tempcntb\m@ne
       \@citea\def\@citea{,}\hbox{\csname b@\@citeb\endcsname}%
     \else
      \advance\@tempcntb\@ne
      \ifnum\@tempcntb=\@tempcntc
      \else\advance\@tempcntb\m@ne\@citeo
      \@tempcnta\@tempcntc\@tempcntb\@tempcntc\fi\fi}}\@citeo}{#1}}
\def\@citeo{\ifnum\@tempcnta>\@tempcntb\else\@citea\def\@citea{,}%
  \ifnum\@tempcnta=\@tempcntb\the\@tempcnta\else
   {\advance\@tempcnta\@ne\ifnum\@tempcnta=\@tempcntb \else \def\@citea{--}\fi
    \advance\@tempcnta\m@ne\the\@tempcnta\@citea\the\@tempcntb}\fi\fi}
 
\makeatother
\begin{titlepage}
\pagenumbering{roman}
\CERNpreprint{\DpPaperGroup}{\DpPaperRef} 
\date{{\small\DpDate}} 
\title{\DpTitle} 
\address{\DpAuthors} 
\begin{shortabs} 
\noindent
A search for pair-produced doubly charged Higgs bosons has been performed 
using the data collected by the DELPHI detector at LEP at centre-of-mass 
energies between 189 and 209 GeV. No excess is observed in the data with 
respect to the Standard Model background. A lower limit for the 
mass of 97.3 GeV/$c^2$ at the 95\% confidence level has been set 
for doubly charged Higgs bosons in left-right symmetric models for 
any value of the Yukawa coupling between the Higgs bosons and the $\tau$ 
leptons.

\vskip 2cm

\centerline
{\it This paper is dedicated to the memory of Paolo Poropat.}
\end{shortabs}
\vfill
\begin{center}
\DpSubmit \ \\ 
\DpComment \ \\
\DpEMail \ \\
\end{center}
\vfill
\clearpage
\headsep 10.0pt
\addtolength{\textheight}{10mm}
\addtolength{\footskip}{-5mm}
\begingroup
%
\newcommand{\DpName}[2]{\hbox{#1$^{\ref{#2}}$},\hfill}
\newcommand{\DpNameTwo}[3]{\hbox{#1$^{\ref{#2},\ref{#3}}$},\hfill}
\newcommand{\DpNameThree}[4]{\hbox{#1$^{\ref{#2},\ref{#3},\ref{#4}}$},\hfill}
\newskip\Bigfill \Bigfill = 0pt plus 1000fill
\newcommand{\DpNameLast}[2]{\hbox{#1$^{\ref{#2}}$}\hspace{\Bigfill}}
%
\footnotesize
\noindent
\DpName{J.Abdallah}{LPNHE}
\DpName{P.Abreu}{LIP}
\DpName{W.Adam}{VIENNA}
\DpName{P.Adzic}{DEMOKRITOS}
\DpName{T.Albrecht}{KARLSRUHE}
\DpName{T.Alderweireld}{AIM}
\DpName{R.Alemany-Fernandez}{CERN}
\DpName{T.Allmendinger}{KARLSRUHE}
\DpName{P.P.Allport}{LIVERPOOL}
\DpName{U.Amaldi}{MILANO2}
% ...
\normalsize
\endgroup
\titlefoot{Department of Physics and Astronomy, Iowa State
     University, Ames IA 50011-3160, USA
    \label{AMES}}
\titlefoot{Physics Department, Universiteit Antwerpen,
     Universiteitsplein 1, B-2610 Antwerpen, Belgium \\
     \indent~~and IIHE, ULB-VUB,
     Pleinlaan 2, B-1050 Brussels, Belgium \\
     \indent~~and Facult\'e des Sciences,
     Univ. de l'Etat Mons, Av. Maistriau 19, B-7000 Mons, Belgium
    \label{AIM}}
\titlefoot{Physics Laboratory, University of Athens, Solonos Str.
     104, GR-10680 Athens, Greece
    \label{ATHENS}}
\titlefoot{Department of Physics, University of Bergen,
     All\'egaten 55, NO-5007 Bergen, Norway
    \label{BERGEN}}
\titlefoot{Dipartimento di Fisica, Universit\`a di Bologna and INFN,
     Via Irnerio 46, IT-40126 Bologna, Italy
    \label{BOLOGNA}}
\titlefoot{Centro Brasileiro de Pesquisas F\'{\i}sicas, rua Xavier Sigaud 150,
     BR-22290 Rio de Janeiro, Brazil \\
     \indent~~and Depto. de F\'{\i}sica, Pont. Univ. Cat\'olica,
     C.P. 38071 BR-22453 Rio de Janeiro, Brazil \\
     \indent~~and Inst. de F\'{\i}sica, Univ. Estadual do Rio de Janeiro,
     rua S\~{a}o Francisco Xavier 524, Rio de Janeiro, Brazil
    \label{BRASIL}}
\titlefoot{Coll\`ege de France, Lab. de Physique Corpusculaire, IN2P3-CNRS,
     FR-75231 Paris Cedex 05, France
    \label{CDF}}
\titlefoot{CERN, CH-1211 Geneva 23, Switzerland
    \label{CERN}}
\titlefoot{Institut de Recherches Subatomiques, IN2P3 - CNRS/ULP - BP20,
     FR-67037 Strasbourg Cedex, France
    \label{CRN}}
\titlefoot{Now at DESY-Zeuthen, Platanenallee 6, D-15735 Zeuthen, Germany
    \label{DESY}}
\titlefoot{Institute of Nuclear Physics, N.C.S.R. Demokritos,
     P.O. Box 60228, GR-15310 Athens, Greece
    \label{DEMOKRITOS}}
\titlefoot{FZU, Inst. of Phys. of the C.A.S. High Energy Physics Division,
     Na Slovance 2, CZ-180 40, Praha 8, Czech Republic
    \label{FZU}}
\titlefoot{Dipartimento di Fisica, Universit\`a di Genova and INFN,
     Via Dodecaneso 33, IT-16146 Genova, Italy
    \label{GENOVA}}
\titlefoot{Institut des Sciences Nucl\'eaires, IN2P3-CNRS, Universit\'e
     de Grenoble 1, FR-38026 Grenoble Cedex, France
    \label{GRENOBLE}}
\titlefoot{Helsinki Institute of Physics, HIP,
     P.O. Box 9, FI-00014 Helsinki, Finland
    \label{HELSINKI}}
\titlefoot{Joint Institute for Nuclear Research, Dubna, Head Post
     Office, P.O. Box 79, RU-101 000 Moscow, Russian Federation
    \label{JINR}}
\titlefoot{Institut f\"ur Experimentelle Kernphysik,
     Universit\"at Karlsruhe, Postfach 6980, DE-76128 Karlsruhe,
     Germany
    \label{KARLSRUHE}}
\titlefoot{Institute of Nuclear Physics,Ul. Kawiory 26a,
     PL-30055 Krakow, Poland
    \label{KRAKOW1}}
\titlefoot{Faculty of Physics and Nuclear Techniques, University of Mining
     and Metallurgy, PL-30055 Krakow, Poland
    \label{KRAKOW2}}
\titlefoot{Universit\'e de Paris-Sud, Lab. de l'Acc\'el\'erateur
     Lin\'eaire, IN2P3-CNRS, B\^{a}t. 200, FR-91405 Orsay Cedex, France
    \label{LAL}}
\titlefoot{School of Physics and Chemistry, University of Lancaster,
     Lancaster LA1 4YB, UK
    \label{LANCASTER}}
\titlefoot{LIP, IST, FCUL - Av. Elias Garcia, 14-$1^{o}$,
     PT-1000 Lisboa Codex, Portugal
    \label{LIP}}
\titlefoot{Department of Physics, University of Liverpool, P.O.
     Box 147, Liverpool L69 3BX, UK
    \label{LIVERPOOL}}
\titlefoot{LPNHE, IN2P3-CNRS, Univ.~Paris VI et VII, Tour 33 (RdC),
     4 place Jussieu, FR-75252 Paris Cedex 05, France
    \label{LPNHE}}
\titlefoot{Department of Physics, University of Lund,
     S\"olvegatan 14, SE-223 63 Lund, Sweden
    \label{LUND}}
\titlefoot{Universit\'e Claude Bernard de Lyon, IPNL, IN2P3-CNRS,
     FR-69622 Villeurbanne Cedex, France
    \label{LYON}}
\titlefoot{Dipartimento di Fisica, Universit\`a di Milano and INFN-MILANO,
     Via Celoria 16, IT-20133 Milan, Italy
    \label{MILANO}}
\titlefoot{Dipartimento di Fisica, Univ. di Milano-Bicocca and
     INFN-MILANO, Piazza della Scienza 2, IT-20126 Milan, Italy
    \label{MILANO2}}
\titlefoot{IPNP of MFF, Charles Univ., Areal MFF,
     V Holesovickach 2, CZ-180 00, Praha 8, Czech Republic
    \label{NC}}
\titlefoot{NIKHEF, Postbus 41882, NL-1009 DB
     Amsterdam, The Netherlands
    \label{NIKHEF}}
\titlefoot{National Technical University, Physics Department,
     Zografou Campus, GR-15773 Athens, Greece
    \label{NTU-ATHENS}}
\titlefoot{Physics Department, University of Oslo, Blindern,
     NO-0316 Oslo, Norway
    \label{OSLO}}
\titlefoot{Dpto. Fisica, Univ. Oviedo, Avda. Calvo Sotelo
     s/n, ES-33007 Oviedo, Spain
    \label{OVIEDO}}
\titlefoot{Department of Physics, University of Oxford,
     Keble Road, Oxford OX1 3RH, UK
    \label{OXFORD}}
\titlefoot{Dipartimento di Fisica, Universit\`a di Padova and
     INFN, Via Marzolo 8, IT-35131 Padua, Italy
    \label{PADOVA}}
\titlefoot{Rutherford Appleton Laboratory, Chilton, Didcot
     OX11 OQX, UK
    \label{RAL}}
\titlefoot{Dipartimento di Fisica, Universit\`a di Roma II and
     INFN, Tor Vergata, IT-00173 Rome, Italy
    \label{ROMA2}}
\titlefoot{Dipartimento di Fisica, Universit\`a di Roma III and
     INFN, Via della Vasca Navale 84, IT-00146 Rome, Italy
    \label{ROMA3}}
\titlefoot{DAPNIA/Service de Physique des Particules,
     CEA-Saclay, FR-91191 Gif-sur-Yvette Cedex, France
    \label{SACLAY}}
\titlefoot{Instituto de Fisica de Cantabria (CSIC-UC), Avda.
     los Castros s/n, ES-39006 Santander, Spain
    \label{SANTANDER}}
\titlefoot{Inst. for High Energy Physics, Serpukov
     P.O. Box 35, Protvino, (Moscow Region), Russian Federation
    \label{SERPUKHOV}}
\titlefoot{J. Stefan Institute, Jamova 39, SI-1000 Ljubljana, Slovenia
     and Laboratory for Astroparticle Physics,\\
     \indent~~Nova Gorica Polytechnic, Kostanjeviska 16a, SI-5000 Nova Gorica, Slovenia, \\
     \indent~~and Department of Physics, University of Ljubljana,
     SI-1000 Ljubljana, Slovenia
    \label{SLOVENIJA}}
\titlefoot{Fysikum, Stockholm University,
     Box 6730, SE-113 85 Stockholm, Sweden
    \label{STOCKHOLM}}
\titlefoot{Dipartimento di Fisica Sperimentale, Universit\`a di
     Torino and INFN, Via P. Giuria 1, IT-10125 Turin, Italy
    \label{TORINO}}
\titlefoot{INFN,Sezione di Torino, and Dipartimento di Fisica Teorica,
     Universit\`a di Torino, Via P. Giuria 1,\\
     \indent~~IT-10125 Turin, Italy
    \label{TORINOTH}}
\titlefoot{Dipartimento di Fisica, Universit\`a di Trieste and
     INFN, Via A. Valerio 2, IT-34127 Trieste, Italy \\
     \indent~~and Istituto di Fisica, Universit\`a di Udine,
     IT-33100 Udine, Italy
    \label{TU}}
\titlefoot{Univ. Federal do Rio de Janeiro, C.P. 68528
     Cidade Univ., Ilha do Fund\~ao
     BR-21945-970 Rio de Janeiro, Brazil
    \label{UFRJ}}
\titlefoot{Department of Radiation Sciences, University of
     Uppsala, P.O. Box 535, SE-751 21 Uppsala, Sweden
    \label{UPPSALA}}
\titlefoot{IFIC, Valencia-CSIC, and D.F.A.M.N., U. de Valencia,
     Avda. Dr. Moliner 50, ES-46100 Burjassot (Valencia), Spain
    \label{VALENCIA}}
\titlefoot{Institut f\"ur Hochenergiephysik, \"Osterr. Akad.
     d. Wissensch., Nikolsdorfergasse 18, AT-1050 Vienna, Austria
    \label{VIENNA}}
\titlefoot{Inst. Nuclear Studies and University of Warsaw, Ul.
     Hoza 69, PL-00681 Warsaw, Poland
    \label{WARSZAWA}}
\titlefoot{Fachbereich Physik, University of Wuppertal, Postfach
     100 127, DE-42097 Wuppertal, Germany \\
\noindent
{$^\dagger$~deceased}
    \label{WUPPERTAL}}

\addtolength{\textheight}{-10mm}
\addtolength{\footskip}{5mm}
\clearpage
\headsep 30.0pt
\end{titlepage}
%
\pagenumbering{arabic} 
\setcounter{footnote}{0} %
\large
\section{Introduction}
Doubly charged Higgs bosons ($H^{\pm\pm}$) appear in several extensions 
to the Standard Model~\cite{hpphmmtheory}, such as
left-right symmetric models, and can be relatively light. 
In Supersymmetric left-right models usually the $SU(2)_R$ gauge symmetry is
broken by two triplet Higgs fields, so-called left and right handed.
 Pair-production of doubly charged 
Higgs bosons is expected to occur mainly via $s$-channel 
exchange of a photon or a \rm{Z} boson. In left-right symmetric models 
the cross-section of $e^+e^- \to H^{++}_L H^{--}_L$ is different from that for 
$e^+e^- \to H^{++}_R H^{--}_R$, where $H^{\pm\pm}_L$ and $H^{\pm\pm}_R$ are the
left-handed and right-handed Higgs bosons. The formulae for the decays 
and the production of these particles can be found in~\cite{hpphmmlimits}.

In these models the doubly charged Higgs boson couples only to charged lepton pairs, 
other Higgs bosons, and gauge bosons, at the tree level. 
The current limit and the mass range of this analysis 
is restricted to the interval between 45 GeV/$c^2$, the
LEP1 limit set by OPAL~\cite{hpphmmz}, and the kinematic limit at LEP2,
that is around 104 GeV/$c^2$. The dominant decay mode of 
the doubly charged Higgs boson is expected to be 
a same sign charged lepton pair, the decay proceeding via a lepton number 
violating coupling. As discussed in~\cite{hpphmmlimits}, due to limits that 
exist for the couplings of 
$H^{\pm\pm} \to e^\pm e^\pm$ from high energy Bhabha scattering, 
$H^{\pm\pm} \to \mu^\pm \mu^\pm$ from the absence of muonium to anti-muonium
transitions and $H^{\pm\pm} \to \mu^\pm e^\pm$ from limits on the flavour
changing decay $\mu^\pm \to e^\mp e^\pm e^\pm$, electron and muon decays 
are not likely. In addition, most of the models expect that the coupling to
$\tau\tau$ will be much larger than any of the others. Therefore, 
only the doubly charged Higgs boson decay $H^{\pm\pm} \to \tau^\pm \tau^\pm$ 
is considered here.

The partial width for the $H^{\pm\pm}$ decay into two $\tau$ leptons 
is, at the tree level~\cite{hpphmmlimits}:

\begin{equation}
\Gamma_{\tau\tau}(H^{\pm\pm} \to \tau^\pm \tau^\pm) = \frac{h^2_{\tau\tau}}
{8\pi} m_H \bigl(1-\frac{2m^2_{\tau}}{m^2_H}\bigr)
\bigl(1-\frac{4m^2_{\tau}}{m^2_H}\bigr)^{1/2}
\end{equation}

\noindent
where $m_\tau$ is the mass of the
$\tau$ lepton and $h_{\tau\tau}$ is the unknown Yukawa coupling
constant. Depending on the $h_{\tau\tau}$ coupling and the Higgs boson 
mass the experimental signature is different. If $h_{\tau\tau}$ is 
sufficiently large, $h_{\tau\tau} \geq 10^{-7}$, 
the Higgs boson decays very close to the interaction point. We describe here an
analysis to search for such events. If $h_{\tau\tau}$ is smaller the decay
occurs inside the tracking detectors or even beyond them, making this analysis
inefficient. In this case pre-existing analyses were applied which 
are further discussed below.
 
\section{Data sample and event generators}
The data collected by DELPHI during the LEP runs at centre-of-mass
energies from 189~GeV to 209~GeV were used. The  total integrated
luminosity of these data samples is $\sim$ 570 pb$^{-1}$. 
The DELPHI detector and its performance have already been described in
detail elsewhere~\cite{delphidet1,delphidet2}.

Signal samples were simulated using the PYTHIA generator~\cite{pythia}. In 
this analysis samples with doubly charged Higgs boson with masses 
between 50 and 100 GeV/$c^2$, in 10 GeV/$c^2$ steps, were used at different 
centre-of-mass energies, both for left-handed and right-handed bosons, and 
different Yukawa coupling constants.

The background estimates from the different Standard Model processes
were based on the following event generators, interfaced with the full 
DELPHI simulation program~\cite{delphidet2}. The 
WPHACT~\cite{wphact} generator was used to produce
four fermion Monte Carlo simulation events. The four fermion samples were 
complemented with dedicated two photon collision samples generated with 
BDK, BDKRC~\cite{bdk} and PYTHIA~\cite{pythia}. Samples of 
$q\bar{q}(\gamma)$ and $\mu^+\mu^-(\gamma)$ events were simulated 
with the KK2f generator~\cite{kk2f}. Finally, KORALZ~\cite{koralz} was 
used to simulate $\tau^+\tau^-(\gamma)$ events and the
generator BHWIDE~\cite{bhwide} was used for $e^+e^-(\gamma)$ events.

\section{Data selection}
The search for pair-produced doubly charged Higgs bosons makes use of three
different analyses depending on the $h_{\tau\tau}$ coupling or, equivalently, on the mean
decay length of the Higgs bosons. When the mean decay length of the 
Higgs boson is very small, the resulting final state 
consists of four narrow and low multiplicity jets coming from the interaction 
point. This analysis is explained in detail in section~\ref{sec:smallpar}. 
For intermediate mean decay lengths of the Higgs boson the topology consists 
of two tracks coming from the interaction point, and with either secondary vertices 
or kinked tracks. If the Higgs boson decays outside the tracking devices the 
signature corresponds to stable heavy massive particles. These two analyses 
were designed for the search for supersymmetric particles decaying to similar
topologies. Details can be found in~\cite{hpphmmstable}.

\subsection{Small impact parameter search}
\label{sec:smallpar}

An initial set of cuts was applied to select events with four jets of low
multiplicity. Only tracks with an impact parameter below 4 cm both in the 
plane transverse to the beam axis and in the direction along the beam axis were
considered in the analysis. 
A charged particle multiplicity between 4 and 8 was required. Events were
clustered into jets using the LUCLUS algorithm~\cite{pythia}, 
requiring each jet to be separated from the others by at least 15
degrees, and only events with four reconstructed jets were accepted. 
To improve the reconstruction of the $\tau$ energy, the
$\tau$ momenta were rescaled, imposing energy
and momentum conservation and keeping the $\tau$ directions at their measured
values. If the rescaled momentum of any jet was negative, the event was rejected, 
as such events are commonly not genuine four jet events.

The two photon background was reduced by the following energy
and momentum requirements: the energy of observed particles produced at a half 
opening angle to the beam axis exceeding $25^\circ$ had to be greater 
than $0.15 \sqrt{s}$, the momenta of the jets 
were required to be larger than $0.01 \sqrt{s}$ and the total neutral energy 
had to be less than $0.35 \sqrt{s}$.

The four lepton background was rejected by requiring that the momentum 
of the most energetic lepton identified (electron or muon) was less than 
$0.25 \sqrt{s}$ and the momentum of the second most energetic lepton identified 
was less than $0.15 \sqrt{s}$. The algorithms used in the lepton identification 
were the same as those used in the selection of fully-leptonic 
$W$ pairs~\cite{wwpaper}.

The calculated $\tau$ momenta, defined above, were used to
reconstruct the Higgs boson mass. The charge of the $\tau$ jet was calculated 
as the sum of the charges of its constituent particles. If this value was not $\pm$1, then the charge
of the most energetic charged particle was assumed to be the charge 
of the $\tau$. For events with two positive $\tau$ lepton
candidates and two negative $\tau$ lepton candidates the charge was used to
assign the pairing of both doubly charged Higgs bosons. If the total charge 
was not equal to 0, the pairing was chosen to minimise the difference 
between the two reconstructed masses of the Higgs bosons. The ratio 
$\frac{|M_{{Rec}^{++}}-M_{{Rec}^{--}}|}{(M_{{Rec}^{++}}+M_{{Rec}^{--}})/2}$ was required to be less
than 0.7. Finally the reconstructed event mass, defined as the average 
of the two masses, had to be greater than 40 GeV/$c^2$.

\begin{figure}[hbtp]
\vspace{1.5cm}
\begin{center}
\epsfig{figure=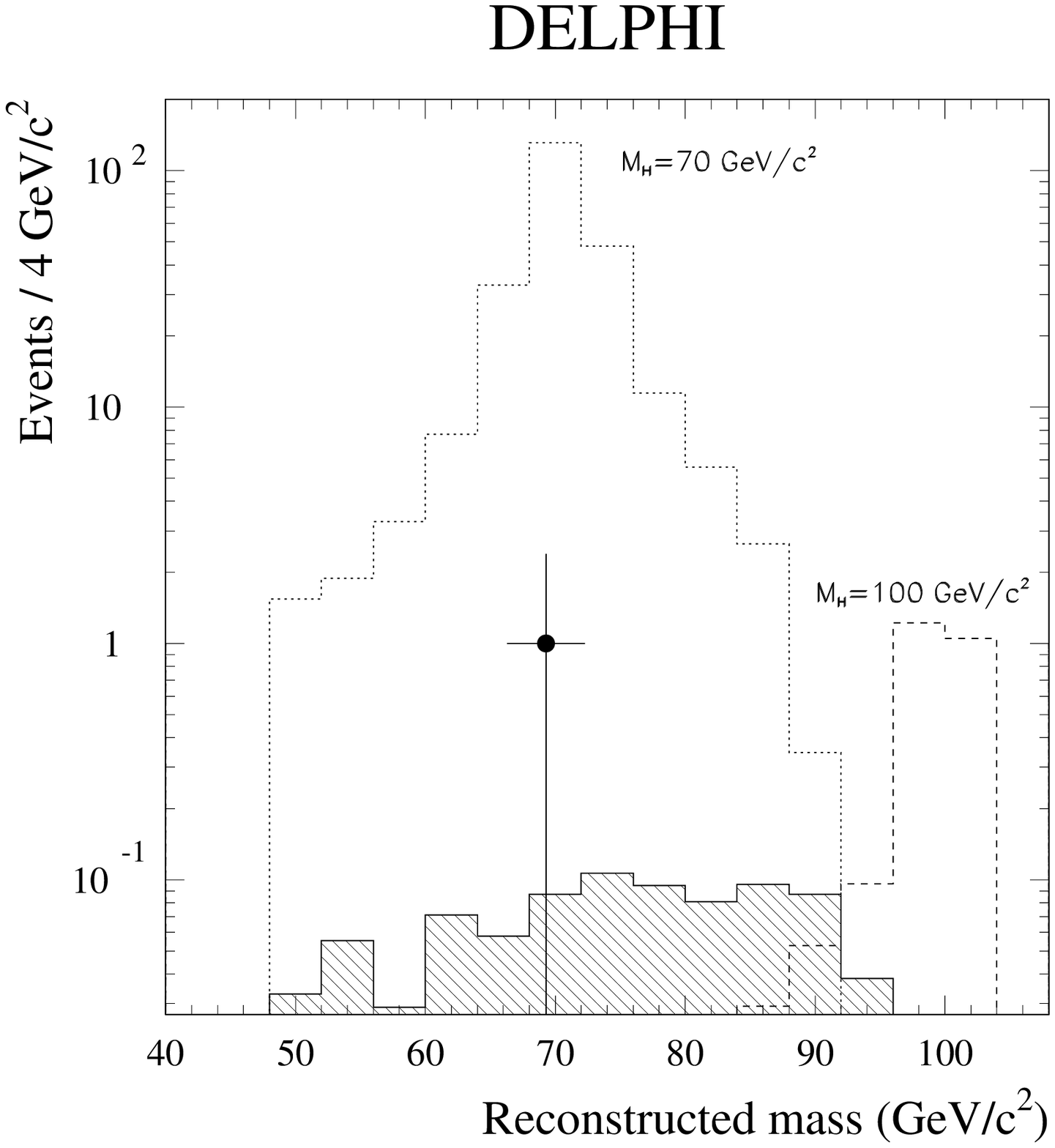,width=15cm,height=15cm}
\end{center}
\caption[]{The reconstructed mass distribution after all cuts 
for the small impact parameter search. The 
hatched histogram corresponds to the expected background and the dot with 
the error bar shows the one remaining candidate event. The 
dotted line corresponds to simulated events with 
$m_{\mathrm{H}^{\pm\pm}_L} = 70$~GeV/$c^2$ and the dashed line 
corresponds to simulated events with $m_{\mathrm{H}^{\pm\pm}_L} = 100$~GeV/$c^2$.}
\label{fig:finalmass}
\end{figure}

\begin{figure}[hbtp]
\vspace{-0.4cm}
\begin{center}
\epsfig{figure=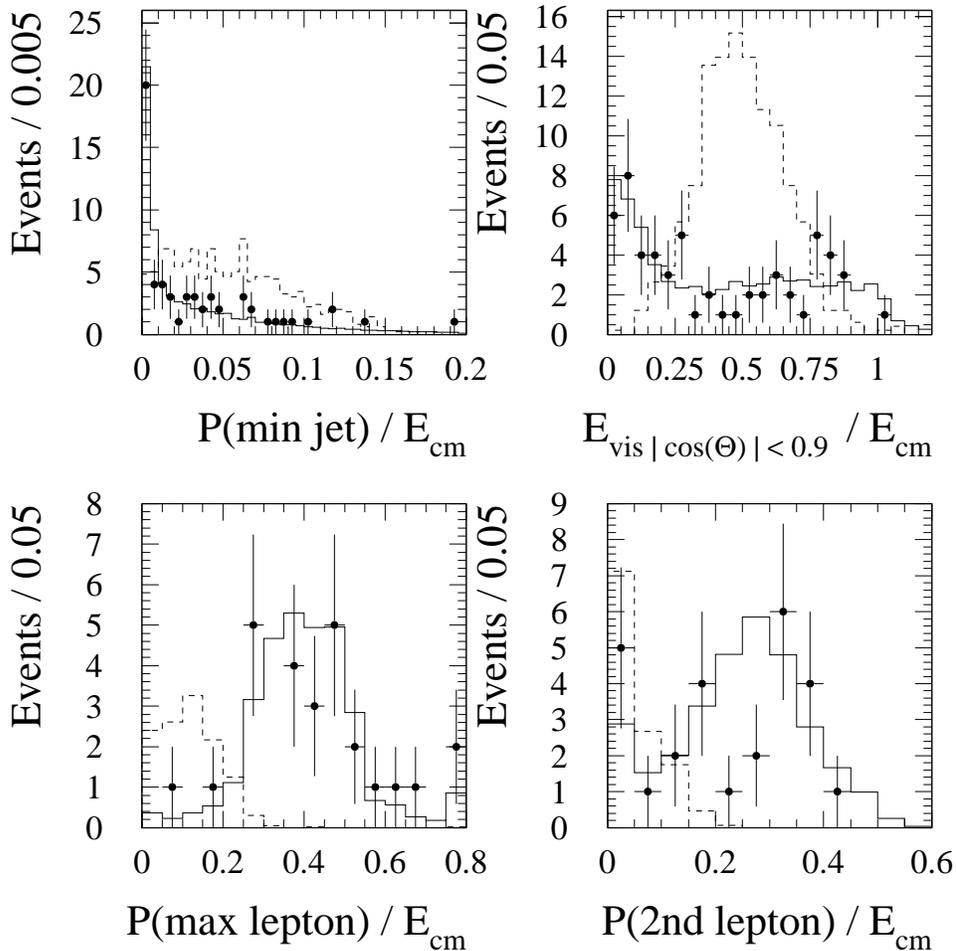,width=15cm,height=15cm}
\end{center}
\caption[]{Event selection variable distributions at different stages of the 
analysis for the small impact parameter search. The top plots show 
the minimum momentum of the jets and the 
visible energy outside $25^\circ$ around the beam axis scaled by $\sqrt{s}$ 
after the four jet preselection cuts. The bottom plots show the momentum 
of the most energetic identified lepton and 
the momentum of the second most energetic identified lepton scaled by $\sqrt{s}$ 
after the anti $\gamma\gamma$ cuts. The
solid lines show the expected background, the dots the observed data and 
the dashed lines correspond to $m_{\mathrm{H}^{\pm\pm}_L} = 100$~GeV/$c^2$. The 
signal is multiplied by a factor 35 in the top plots and by a factor 4 in the
bottom plots.}
\label{fig:preselcuts}
\end{figure}

The effects of the selection cuts are shown in Table~\ref{tab:cutevol} 
for the combined 189-209 GeV sample. After all cuts were applied only one 
event was observed in the data with a mass of 69$\pm$3 GeV/c$^2$, 
while 0.9 events were expected from background 
processes. The candidate was collected at $\sqrt{s}$=206.7 GeV and 
is compatible with the assignment $ZZ \to \tau^+ \tau^- \tau^+ \tau^-$. The most probable
reconstructed masses with different sign leptons are indeed 
compatible with a $M_Z$-$M_Z$ mass hypothesis at the one sigma level. The signal 
efficiency was around 40\% for a wide range of masses between 70 
and 100 GeV/$c^2$ 
for both left-handed and right-handed doubly charged Higgs bosons, as shown in 
Table~\ref{tab:effsmall}. Table~\ref{tab:effhtt} shows the selection 
efficiencies for left-handed doubly charged Higgs bosons for several 
$H^{\pm\pm}$ masses and several $h_{\tau\tau}$ couplings at 
$\sqrt{s}$=206.7 GeV. 
The final reconstructed mass spectrum and the expected mass 
distribution in simulated signal events are shown in Figure~\ref{fig:finalmass}. 
The good level of agreement between data and simulation observed at different 
stages of the analysis is demonstrated in Figure~\ref{fig:preselcuts}.


\begin{table} [h!]
\begin{center}
\vspace{1ex}
\begin{tabular}{lrrrrr}
\hline
cut & \multicolumn{1}{c}{data}
    & \multicolumn{1}{c}{total bkg.}
    & \multicolumn{1}{c}{$llll$}
    & \multicolumn{1}{c}{other}
    & \multicolumn{1}{c}{$\varepsilon_{H^{++}_L H^{--}_L}$} \\
\hline
Four jets preselection     &  59 &  67.41$\pm$0.95    &  44.01$\pm$0.31 & 23.40$\pm$0.90   &   59.2\% \\
anti $\gamma\gamma$ cuts  &   26 &  31.03$\pm$0.48    &  28.90$\pm$0.25 &  2.13$\pm$0.41   &   52.3\% \\
anti 4 lepton cuts        &    1 &   1.87$\pm$0.07    &   1.69$\pm$0.06 &  0.18$\pm$0.03   &   48.7\% \\
Mass requirements         &    1 &   0.91$\pm$0.04    &   0.85$\pm$0.04 &  0.06$\pm$0.01   &   44.2\% \\
\hline
\end{tabular}
\end{center}
\caption{The total number of events observed and the expected background 
after the different cuts used in the analysis for the small impact parameter 
search for the combined 189-209 GeV sample. The errors are only statistical. 
The last column shows the efficiency for a left-handed doubly charged Higgs 
boson signal with $m_{\mathrm{H}^{\pm\pm}_L} = 100$~GeV/$c^2$ at 
$\sqrt{s}$=206.7 GeV. The statistical error in the signal efficiency is 
about 1.5\% in all cases.}
\label{tab:cutevol}
\end{table}

\begin{table}[htbp]
\begin{center}
\begin{tabular}{|c|c|c|c|c|c|c|}
\hline
channel & \multicolumn{6}{c|}
    {$M_{H^{\pm\pm}}$ (GeV/c$^2$)  } \\ \cline{2-7}
        &  50 & 60 & 70 & 80 & 90 & 100 \\
\hline
\hline
left-handed   & 32.7 & 36.6 & 40.5 & 44.8  & 43.4  & 44.2  \\
right-handed  & 31.8 & 37.0 & 40.0 & 44.0  & 44.8  & 45.2  \\
\hline
\end{tabular}
\end{center}
\caption{Selection efficiencies (in \%) for left-handed and right-handed 
$H^{++}H^{--} \to$ $\tau^+\tau^+\tau^-\tau^-$ for several $H^{\pm\pm}$ 
masses and $h_{\tau\tau} \geq 10^{-7}$ at $\sqrt{s}$=206.7 GeV, 
for the small impact parameter search. The statistical error is 
around 1.5\% in all cases.}
\label{tab:effsmall}
\end{table}

\begin{table}[htbp]
\begin{center}
\begin{tabular}{|c|c|c|c|c|c|c|}
\hline
$h_{\tau\tau}$ & \multicolumn{4}{c|}
    {$M_{H^{\pm\pm}}$ (GeV/c$^2$)  } \\ \cline{2-5}
        &  50 &  70 &  90 & 100 \\
\hline
\hline
 $ 4 \cdot 10^{-8}$   & 0.2/38.1/13.1  &  1.6/43.0/1.4 &    6.0/23.9/0.0  &  20.5/5.3/0.0  \\
\hline
 $ 10^{-8}$           & 0.0/6.4/68.4   & 0.0/16.0/57.2 &   0.0/30.5/22.7  &  0.0/36.3/7.3  \\
\hline
 $  \leq 10^{-9} $    & 0.0/0.0/77.6   &  0.0/0.0/77.6 &   0.0/0.0/41.3   &  0.0/0.0/41.6  \\
\hline
\end{tabular}
\end{center}
\caption{Selection efficiencies (in \%) for left-handed doubly charged Higgs
bosons for several $H^{\pm\pm}$ masses and several $h_{\tau\tau}$ couplings 
at $\sqrt{s}$=206.7 GeV, for the three analyses performed 
(small impact parameter search, search for secondary vertices or kinks and 
search for stable massive particles, respectively). The statistical error 
is around 1.5\% in all cases.}
\label{tab:effhtt}
\end{table}

\subsubsection{Systematic uncertainties}

Several sources of systematic uncertainties on the signal efficiency and 
the background level were investigated. The
particle identification was checked on di-lepton samples both at the Z peak 
and at high energy. The discrepancy in the efficiencies between the data and the 
simulation was found to be lower than 2\% in all cases. The 
track selection and the track reconstruction efficiency was also studied with 
these samples. These effects were studied by the comparison between data and 
simulation for tracks at the boundaries of subdetector acceptances, where 
systematic effects are expected to be larger. The systematic error of these 
effects was about 1.5\%.

\begin{figure}[hbtp]
\vspace{-0.4cm}
\begin{center}
\epsfig{figure=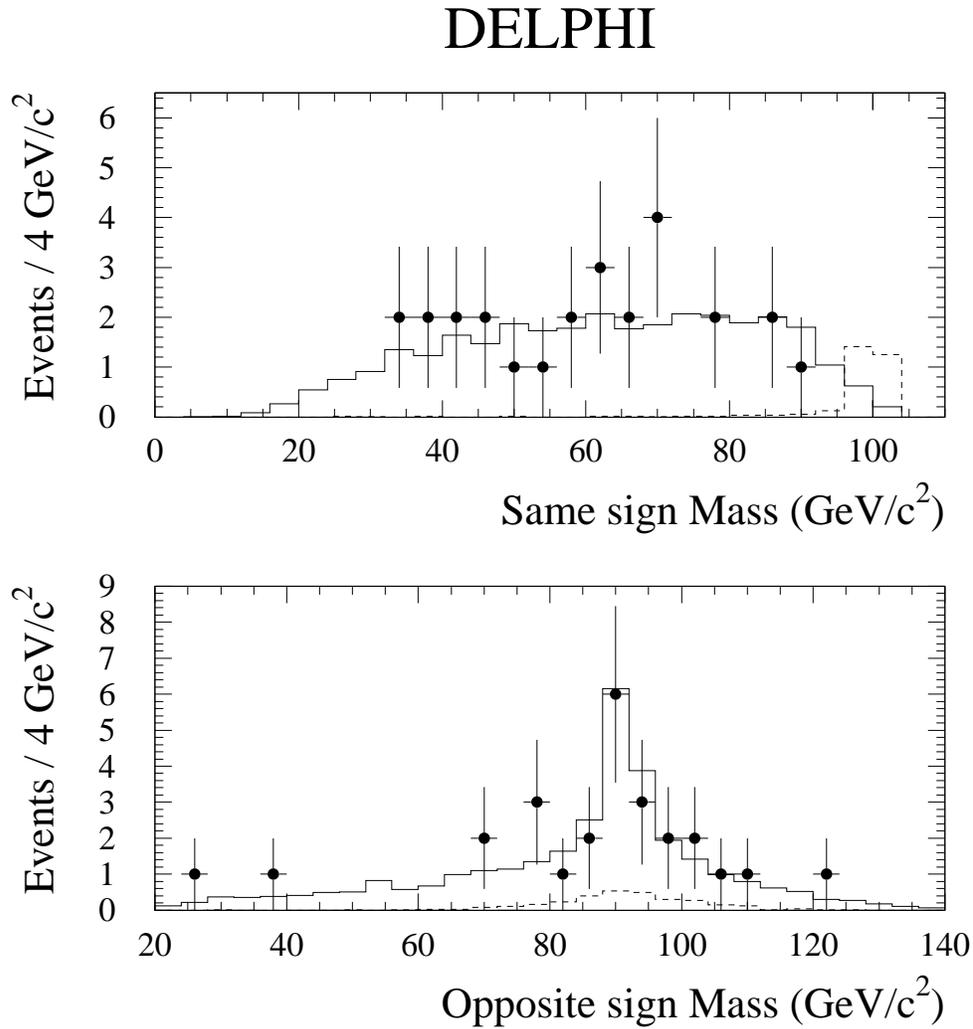,width=15cm,height=15cm}
\end{center}
\caption[]{Reconstructed mass distributions for the small impact
parameter search. The masses are shown for the same sign lepton pairs
(top) and the opposite sign lepton combination closest to the Z mass
(bottom). These distributions are shown before the anti four lepton cuts.
The solid lines show the expected background, the dots the observed
data and 
the dashed lines correspond to simulated events with 
$m_{\mathrm{H}^{\pm\pm}_L} = 100$~GeV/$c^2$.}
\label{fig:mass_presel}
\end{figure}

The errors on the background and signal rates from the modelling of 
the detector response were a few percent. 
Different variables at preselection level have been
studied, with good agreement between data and simulation observed. 
The distributions in relevant variables before the anti $\gamma\gamma$ cuts 
and the anti four lepton cuts are shown 
in Figure~\ref{fig:preselcuts}. The masses reconstructed from both
same sign and different sign lepton pairs, before the anti four lepton cuts 
were applied, are shown 
in Figure~\ref{fig:mass_presel}. For the opposite sign lepton pairs only the 
mass of the combination closest to the \rm{Z} mass has been given and the \rm{Z} 
peak is visible.

The total systematic error on the background was about 13\%, with a dominant 
contribution of about 12\% due to the limited simulation statistics available. 
The total systematic error on the efficiency was about 5\%.

\subsection{Search for secondary vertices or kinks}
When the lifetime is such that the particle decays inside the tracking detector,
the previous analysis is inefficient, because impact parameter cuts are applied
to reject the background coming from secondary interactions. We have applied
here the analysis described in~\cite{hpphmmstable}, that performs a special 
track reconstruction for this particular topology, looking for decay 
vertices far from the interaction point.

After all cuts five events were selected in the data, while 2.9 events were
expected from the background. The signal efficiency was about 40\%, 
if the mean decay length was about 50~$\rm{cm}$ with a smooth fall for both lower and
higher mean decay lengths. The selection efficiencies for several 
$H^{\pm\pm}$ masses and several $h_{\tau\tau}$ couplings at 
$\sqrt{s}$=206.7 GeV are shown in Table~\ref{tab:effhtt}.

\subsection{Search for stable massive particles}
If the lifetime is even larger, the $H^{\pm\pm}$ crosses the tracking devices 
without decaying. The analysis described in~\cite{hpphmmstable} to search for stable
heavy particles is applied here. It is based on the measurement of anomalous 
ionisation loss measured in the Time Projection Chamber and of the absence of 
Cherenkov light detected in the Ring Imaging Cherenkov Detector.

One event was selected in the data, in agreement with the expected background 
of 1.9 events. For stable particle masses in the range of 50-80 GeV/$c^2$ 
the efficiency was $\sim$ 75\%, decreasing to $\sim$40\% for masses 
near the kinematic limit (Table~\ref{tab:effhtt}).

\section{Determination of the mass limit}

No evidence for $H^{++}H^{--}$ production was observed. A modified 
frequentist likelihood ratio method~\cite{alex} has been used 
to compute the cross-section and mass
limits. The reconstructed event mass was used as a discriminant variable 
in the computation of the confidence levels in the small impact parameter 
analysis, while for the others only the number of events were used. The
systematic errors were taken into account in the computation. All 
centre-of-mass energies and the three analyses were treated as 
independent experiments. For intermediate mean decay lengths of the 
Higgs bosons in many cases two analyses have 
significant efficiency. However the overlap of the samples selected by 
the analyses, both for the signal and for the background, was negligible.

A very similar behaviour, both in terms of efficiency and of mass 
distributions, was observed for the left-handed and the right-handed doubly charged 
Higgs bosons. Hence, the average of both 
contributions were used to calculate the confidence
levels. The expected left-handed and right-handed cross-sections were 
calculated using the PYTHIA generator~\cite{pythia}.

\begin{figure}[hbtp]
\vspace{-0.4cm}
\begin{center}
\epsfig{figure=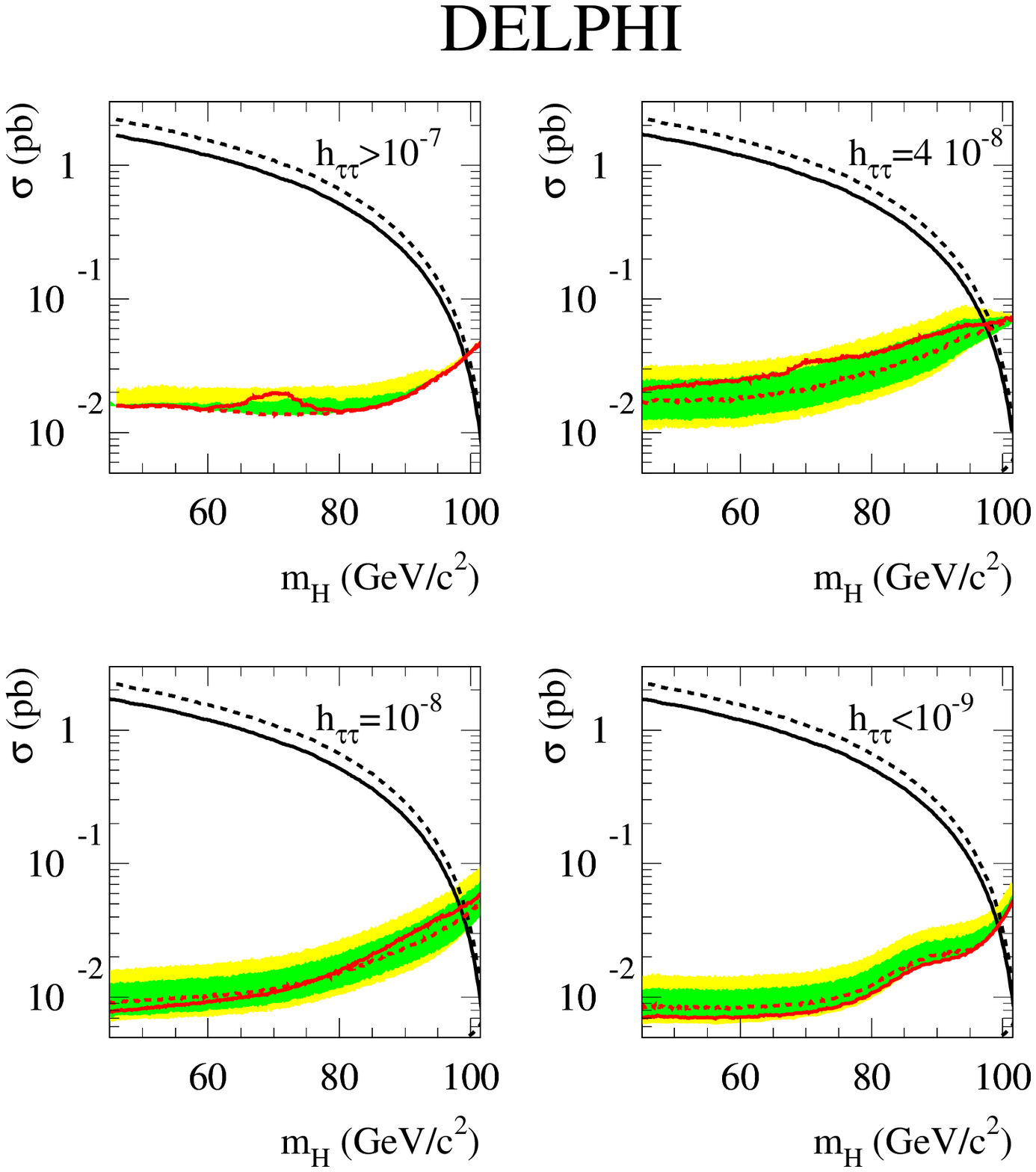,width=15cm,height=15cm}
\end{center}
\caption[]{ 
Upper limits, at 95\% confidence level, on the production cross-section 
for a pair of doubly charged Higgs bosons as a function of the 
doubly charged Higgs boson mass at $\sqrt{s}$=206.7 GeV, 
assuming 100\% branching ratio for the decay of $H^{\pm \pm}$ into 
$\tau^\pm\tau^\pm$ for different values of the $h_{\tau\tau}$ coupling. 
The dashed grey curve shows the expected upper limit with 
one and two standard deviation bands and the solid grey curve is the observed 
upper limit of the cross-section (the grey curves are those inside the bands). 
The dashed black and solid black curves show the expected production 
cross-section of $H^{\pm \pm}_L$ and $H^{\pm \pm}_R$ pairs in 
left-right symmetric models.}
\label{fig:xslimit}
\end{figure}

Previous searches for $H^{\pm\pm}$ pair production have already excluded 
$M_{H^{\pm\pm}} < $ 45.6 GeV/$c^2$~\cite{hpphmmz}. Therefore, this search 
was limited to masses greater than this value. 
The limits at 95\% confidence level for different values of $h_{\tau\tau}$ 
are shown in table~\ref{tab:limits}. 
Figure~\ref{fig:xslimit} shows the 95\% confidence level upper limits on 
the cross-section at $\sqrt{s}$ = 206.7 GeV for
the production of $H^{++}H^{--} \to \tau^+\tau^+\tau^-\tau^-$ for these 
values of $h_{\tau\tau}$. The comparison 
of these limits with the expected cross-section for left-handed 
$H^{\pm \pm}_L$ and right-handed $H^{\pm \pm}_R$ pair production yields 
95\% confidence level lower limits 
on the mass of the $H^{\pm\pm}_L$ and $H^{\pm\pm}_R$ bosons of 98.1 
and 97.3 GeV/$c^2$, respectively for any value of the $h_{\tau\tau}$ coupling.

This search slightly improves previous searches for 
$h_{\tau\tau} \geq 10^{-7}$~\cite{hpphmmopal}, and in addition is extended to the
whole range of the $h_{\tau\tau}$ coupling.

\begin{table}[htbp]
\begin{center}
\begin{tabular}{|c|c|c|c|c|}
\hline
$h_{\tau\tau}$ & \multicolumn{2}{|c|} {Left-handed} & 
\multicolumn{2}{|c|} {Right-handed} \\
        &  Observed & Expected & Observed & Expected \\
\hline
\hline
 $ \geq 10^{-7}$         & 99.6 & 99.6  & 99.1 & 99.1 \\
\hline
 $ 4 \cdot 10^{-8}$  & 98.1 & 98.4  & 97.3 & 97.6  \\
\hline
 $ 10^{-8}$          & 99.0 & 99.4  & 98.4 & 98.9  \\
\hline
 $ \leq 10^{-9}$        & 99.6 & 99.6  & 99.3 & 99.3  \\
\hline
\end{tabular}
\end{center}
\caption{Median expected and observed $H^{\pm\pm}$ mass limits at 95\% C.L. 
in GeV/$c^2$ for different values of the $h_{\tau\tau}$ coupling.}
\label{tab:limits}
\end{table}

\section{Conclusion}
A search for pair-produced doubly charged Higgs bosons decaying into $\tau$ 
leptons was performed using the data collected by DELPHI at LEP at centre-of-mass
energies from 189~GeV to 209~GeV in R-parity conserving supersymmetric
left-right symmetric models. Three different analyses were applied to cover the
whole range of the $h_{\tau\tau}$ coupling: decays very close to 
the interaction point, inside the tracking detectors or beyond them. 
No significant excess was observed and a lower limit on the doubly charged
Higgs boson mass of 97.3~GeV/$c^2$ has been set at 95\% confidence level 
for any value of the $h_{\tau\tau}$ coupling.

\subsection*{Acknowledgements}
\vskip 3 mm
 We are greatly indebted to our technical 
collaborators, to the members of the CERN-SL Division for the excellent 
performance of the LEP collider, and to the funding agencies for their
support in building and operating the DELPHI detector.\\
We acknowledge in particular the support of \\
Austrian Federal Ministry of Education, Science and Culture,
GZ 616.364/2-III/2a/98, \\
FNRS--FWO, Flanders Institute to encourage scientific and technological 
research in the industry (IWT), Belgium,  \\
FINEP, CNPq, CAPES, FUJB and FAPERJ, Brazil, \\
Czech Ministry of Industry and Trade, GA CR 202/99/1362,\\
Commission of the European Communities (DG XII), \\
Direction des Sciences de la Mati$\grave{\mbox{\rm e}}$re, CEA, France, \\
Bundesministerium f$\ddot{\mbox{\rm u}}$r Bildung, Wissenschaft, Forschung 
und Technologie, Germany,\\
General Secretariat for Research and Technology, Greece, \\
National Science Foundation (NWO) and Foundation for Research on Matter (FOM),
The Netherlands, \\
Norwegian Research Council,  \\
State Committee for Scientific Research, Poland, SPUB-M/CERN/PO3/DZ296/2000,
SPUB-M/CERN/PO3/DZ297/2000, 2P03B 104 19 and 2P03B 69 23(2002-2004)\\
JNICT--Junta Nacional de Investiga\c{c}\~{a}o Cient\'{\i}fica 
e Tecnol$\acute{\mbox{\rm o}}$gica, Portugal, \\
Vedecka grantova agentura MS SR, Slovakia, Nr. 95/5195/134, \\
Ministry of Science and Technology of the Republic of Slovenia, \\
CICYT, Spain, AEN99-0950 and AEN99-0761,  \\
The Swedish Natural Science Research Council,      \\
Particle Physics and Astronomy Research Council, UK, \\
Department of Energy, USA, DE-FG02-01ER41155.

\clearpage
\newpage

\begin{thebibliography}{99}

\bibitem{hpphmmtheory} J.F. Gunion, H.E. Haber, G.L. Kane and D. Dawson, The 
Higgs Hunter's Guide, Frontiers in Physics, Lecture Notes Series, 
Addison Wesley, 1990; \\
C. S. Aulakh, A. Melfo and G. Senjanovic, Phys. Rev. {\bf D57} (1998) 4174; \\
Z. Chacko and R. N. Mohapatra, Phys. Rev. {\bf D58} (1998) 15003.
                      
\bibitem{hpphmmlimits} M. L. Swartz, Phys. Rev. {\bf D40} (1989) 1521; \\
                       K. Huitu et al., Nucl. Phys., {\bf B487} (1997) 27.

\bibitem{hpphmmz} OPAL Collaboration, P. D. Acton et al., Phys. 
Lett. {\bf B295} (1992) 347.

\bibitem{delphidet1}
 DELPHI Collaboration, P. Aarnio {\it et al.}, Nucl. Instr. and Meth.
{\bf A 303} (1991) 233.

\bibitem{delphidet2}
DELPHI Collaboration, P. Abreu {\it et al.}, Nucl. Instr. and Meth.
{\bf A 378} (1996) 57.

\bibitem{pythia} T.~Sj\"ostrand, {\it PYTHIA 5.719 / JETSET 7.4},
   Physics at LEP2, eds. G.~Altarelli,
   T.~Sj\"ostrand and F.~Zwirner, CERN 96-01 (1996) Vol 2, 41.

\bibitem{wphact} E. Accomando and A. Ballestrero Comput.Phys.Commun. {\bf 99} 
(1997) 270; \\
E. Accomando, A. Ballestrero and E. Maina hep-ph/0204052 
(2002),  to appear in Comput. Phys. Commun.

\bibitem{bdk} F.A. Berends, P.H. Daverveldt, R. Kleiss, Comp. Phys.
  Comm. {\bf 40} (1986) 271, 285 and 309.  

\bibitem{kk2f} S.~Jadach, B.F.L.~Ward, Z.~Was, Phys. Lett. {\bf B449} (1999) 97; \\
S.~Jadach, B.F.L.~Ward, Z.~Was, Comp. Phys. Comm. {\bf 130} (2000) 260.

\bibitem{koralz} S.~Jadach, B.F.L.~Ward, Z.~Was, Phys. Lett. {\bf 390} 
(1997) 298.

\bibitem{bhwide} S.~Jadach, W. Placzek, B.F.L.~Ward, Comp.
Phys. Comm. {\bf 79} (1994) 503.

\bibitem{hpphmmstable} DELPHI Collaboration, J. Abdallah {\it et al.}, 
Search for supersymmetric particles in light gravitino scenarios, 
CERN-EP 2002-083, Submitted to Eur. Phys. J. C. 

\bibitem{wwpaper} DELPHI Collaboration, P. Abreu {\it et al.}, 
Phys. Lett. {\bf B479} (2000) 89.

\bibitem{alex} A.L. Read, in CERN Yellow Report 2000-005, p.~81.

\bibitem{hpphmmopal} OPAL Collaboration, G. Abbiendi et al., 
Phys. Lett. {\bf B526} (2002) 221-232.

\end{thebibliography}
\end{document}